\documentclass[aps,prd,twocolumn,showkeys,showpacs,superscriptaddress,10pt]{revtex4-1}
\usepackage{graphicx}  % needed for figures
\usepackage{dcolumn}   % needed for some tables
\usepackage{bm,braket,soul,color}        % for math
\usepackage{amssymb,MnSymbol}   % for math
\usepackage{hyperref}
\usepackage{comment}
\usepackage{color}

\pdfoutput=1

\renewcommand{\[}{\begin{equation}}
\renewcommand{\]}{\end{equation}}
\newcommand{\ov}[1]{\overline#1}
\newcommand{\tr}{\mathrm{tr}}
\renewcommand{\det}[1]{|#1|}
\newcommand{\norm}[1]{\left\lvert#1\right\rvert}

\newcommand{\A}{\alpha}
\newcommand{\B}{\beta}
\newcommand{\de}{d\epsilon}
\newcommand{\bd}{\boldsymbol{d}}

\begin{document}

\title{Quantum parameter estimation using multi-mode Gaussian states}

\author{Dominik \v{S}afr\'{a}nek}\email{pmxdd@nottingham.ac.uk}
\affiliation{School of Mathematical Sciences, University of Nottingham, University Park,
Nottingham NG7 2RD, United Kingdom}
\affiliation{Faculty of Physics, University of Vienna, Boltzmanngasse 5, 1090 Vienna, Austria}
\author{Antony R. Lee}
\affiliation{School of Mathematical Sciences, University of Nottingham, University Park,
Nottingham NG7 2RD, United Kingdom}
\author{Ivette Fuentes}\thanks{Previously known as Fuentes-Guridi and Fuentes-Sch\"uller.}
\affiliation{School of Mathematical Sciences, University of Nottingham, University Park,
Nottingham NG7 2RD, United Kingdom}
\affiliation{Faculty of Physics, University of Vienna, Boltzmanngasse 5, 1090 Vienna, Austria}

\date{\today}

\begin{abstract}
Gaussian states are of increasing interest in the estimation of physical parameters because they are easy to prepare and manipulate in experiments. In this article, we derive formulae for the optimal estimation of parameters using two- and multi-mode Gaussian states. As an application of our result, we derive the optimal Gaussian probe states for the estimation of the parameter characterizing a one-mode squeezing channel.
\end{abstract}

\pacs{03.67.-a, 06.20.-f, 03.65.Ta}
\keywords{Quantum metrology, Gaussian states, Local estimation theory}

\maketitle

\section{Introduction}

One of the main aims of quantum metrology is to find the ultimate precision bound on the estimation of a physical parameter encoded in a quantum state. Of special interest are parameters that cannot be measured directly, since they do not correspond to observables of the system. However, they can be estimated by finding an appropriate measurement strategy. The estimation also involves choosing an estimator $\hat{\epsilon}$ which maps the set of the measurement results onto the set of possible parameters. The ultimate precision limit is given by the quantum Cram\'er-Rao bound~\cite{BraunsteinCaves1994a,Paris2009a} which gives a lower bound on the mean squared error of any locally unbiased estimator $\hat{\epsilon}$. The local unbiasedness means that in the limit where the number of measurements goes to infinity, the value of the estimator converges to the real value of the parameter. The bound is given by the number of measurements taken on the identical copies of the state $\hat{\rho}(\epsilon)$ and a quantity $H(\epsilon)$ called the quantum Fisher information. Higher precision is achieved by increasing the number of measurements and maximizing the quantum Fisher information. Calculating the quantum Fisher information thus gives us an idea of how well we can estimate the parameter when only a fixed amount of measurements are available. This technique has been applied, for example, in large interferometers like VIRGO~\cite{Caron1995a} and LIGO~\cite{Abbott2004a} assigned to measure gravitational waves, or a current proposal~\cite{Sabin2014a} of measuring gravitation waves using phonons in Bose-Einstein condensates, magnetometers~\cite{Nusran2014a,Petersen2005a}, and gravimeters~\cite{Bruschi2014a}.

Calculating the quantum Fisher information is not always an easy task. Although a general formula for the quantum Fisher information exists, it is written in a terms of the density matrix~\cite{Paris2009a}. On the other hand, many applications use a special kind of continuous-variable systems called \emph{Gaussian states}, for which the description using density matrices seems particularly ineffective. Gaussian states can be conveniently described in terms of the first and the second moments of the so-called quadrature operators. This description is usually called the phase-space or the covariance matrix formalism~\cite{Adesso2014a}.

Despite of the importance and practical usage of the quantum Fisher information, the theory for estimation using Gaussian states in the phase-space formalism is far from complete, and only partial results are known. The first leap in deriving general formulae has been taken by Pinel et al.~\cite{Pinel2012a}, who found a formula for pure states, i.e., for the states which are pure at point $\epsilon$ and remain pure even if the $\epsilon$ slightly changes. The same year Marian and Marian found the formula for the fidelity between one-mode and two-mode Gaussian states~\cite{Marian2012a}, which allowed for the derivation of the general formula for the one-mode state~\cite{Pinel2013b}. Also, Spedalieri et al.~found a formula for the fidelity between one pure and one mixed Gaussian state~\cite{Spedalieri2013a}, from which one can derive a slightly more general formula for pure states, i.e., for the states which are pure at the point $\epsilon$ but a small change in $\epsilon$ introduces impurity. A different path has been followed by Monras~\cite{Monras2013a}, who connected the quantum Fisher information to the solution of the so-called Stein-equation. Using this approach, he derived the quantum Fisher information for a generalization of pure states called iso-thermal states, and a general formula for any multi-mode Gaussian state in terms of an infinite series. Using the previous result, Jiang derived a formula~\cite{Jiang2014a} for Gaussian states in exponential form and simplified a known formula for pure states. Quite recently, Gao and Lee derived an exact formula~\cite{Gao2014a} for the quantum Fisher information for the multi-mode Gaussian states in terms of the inverse of certain tensor products, elegantly generalizing the previous results, however with some possible drawbacks, especially in the necessity of inverting relatively large matrices.

In this article, we first introduce a phase-space description of the Gaussian states. Then we derive an exact formula for two-mode Gaussian states in analogy with~\cite{Pinel2013b}. We simplify the result of Monras for multi-mode Gaussian states and we show that the infinite sum involved converges as a geometrical series. However, the series may not be easy to evaluate, and for that reason we calculate the error when only a finite number of terms of the sum are taken into account. On the other hand, in the case when the Williamson decomposition of the covariance matrix is known, the infinite series can be evaluated. This gives a general and exact formula for the quantum Fisher information of any multi-mode Gaussian state in terms of its decomposition. Finally, we use the derived formula to find the optimal Gaussian states for estimating the squeezing parameter of a single mode squeezing channel, and we demonstrate that this estimation can be exponentially enhanced by the initial squeezing when followed by an appropriate rotation. In~\ref{app:real_covariance} we connect the so-called real and the complex form of the covariance matrix, in~\ref{app:pure_regularization} we study the case of pure states and the regularization procedure which allows us to use our results for generally mixed states, even for states where some or all modes remain pure. \ref{app:Remainder} and \ref{app:exact_multimode} contain detailed proofs for some results and in~\ref{app:example} we show a full derivation of our single mode squeezing channel example. In~\ref{app:notations} we provide a table of frequently used notation that are constant throughout the paper and appear repeatedly.

\subsection{Phase-space formalism of Bosonic modes and the Williamson decomposition}

In this section we recapitulate the phase-space description of a Bosonic system, which will be particularly useful for the continuous parameter states known as \emph{Gaussian states}. First, let us consider a Bosonic system with the set of annihilation and creation operators $\{\hat{a}_n,\hat{a}_n^\dag\}$. We collect them into a vector of operators $\boldsymbol{\hat{A}}:=(\hat{a}_1,\hat{a}_2,\dots,\hat{a}_1^\dag,\hat{a}_2^\dag,\dots)^T$. Now we can write the commutation relation between these operators in a compact form
\[\label{def:commutation_relation}
[\boldsymbol{\hat{A}}_{i},\boldsymbol{\hat{A}}_{j}^{\dag}]=K_{ij}\mathrm{id}\quad\!\Rightarrow\quad\! K=
\begin{bmatrix}
I & 0 \\
0 & -I
\end{bmatrix},
\]
where $\mathrm{id}$ denotes the identity element of an algebra and $I$ is the identity matrix. Note that $K^{-1}=K^\dag=K$ and $K^2=I$.

One way of representing a state in quantum mechanics is by using a density matrix $\hat{\rho}$, however, for Bosonic systems an alternative and completely equivalent description exists, which is particularly useful in a description of Gaussian states. Given a state $\hat{\rho}$ we define \emph{the symmetric characteristic function} as
\[
\chi(\boldsymbol{\xi})\,=\,\mathrm{tr}[\hat{\rho}\,\hat{D}(\boldsymbol{\xi})],
\]
where $\hat{D}(\boldsymbol{\xi})\,=\,e^{\boldsymbol{\hat{A}}^{\dag}K\boldsymbol{\xi}}$ is the \emph{Weyl displacement operator} with a variable of the form $\boldsymbol{\xi}\,=\,\boldsymbol{\gamma}\oplus\overline{\boldsymbol{\gamma}}$. Gaussian states are those whose characteristic function is, by definition, of Gaussian form, i.e.,
\[
\chi(\boldsymbol{\xi})\,=\,e^{-\frac{1}{4}\boldsymbol{\xi}^{\dag}\sigma\boldsymbol{\xi}-i\,{\bd}^{\dag}K\boldsymbol{\xi}}.
\]
In analogy with classical probability theory, Gaussian states are completely described by the first and the second statistical moments $\bd$ and $\sigma$, where the vector $\bd$ and the positive-definite Hermitian matrix $\sigma$ are defined as
\begin{subequations}\label{def:covariance_matrix}
\begin{align}
\bd_i&=\mathrm{tr}\big[\hat{\rho}\boldsymbol{\hat{A}}_i\big],\\
\sigma_{ij}&=\mathrm{tr}\big[\hat{\rho}\,\{\Delta\boldsymbol{\hat{A}}_i,\Delta\boldsymbol{\hat{A}}^{\dag}_j\}\big],
\end{align}
\end{subequations}
where $\Delta\boldsymbol{\hat{A}}:=\boldsymbol{\hat{A}}-\bd$, and $\{\cdot,\!\cdot\}$ denotes the anti-commutator. We call $\bd$ the displacement vector and $\sigma$ the covariance matrix. Note that other authors use different conventions. We choose the covariance matrix defined by the anti-commutator of annihilation and creation operators which is known as the `complex form', while some authors define it using the correlations between position and momenta operators. In our convention the vacuum is represented by the identity matrix $I$, i.e., the variance of the quadrature operators $\hat{x}_{i}$ and $\hat{p}_{i}$ are $\mathrm{var}\,(\hat{x}_{i})\,=\,\mathrm{var}\,(\hat{p}_{i})\,=\,+1$ (some authors define the vacuum variances as $+1/2$). This is of course only a definition and does not affect any physical interpretation of the results.

According to Williamson's theorem~\cite{Williamson1936a,Simon1998a,Arvind1995a} any positive-definite $2N\times2N$ Hermitian matrix $\sigma$ can be diagonalized using symplectic matrices, i.e. $\sigma=SDS^\dag$, where $S$ is an element of a complex representation~\cite{Arvind1995a} of the real symplectic group $Sp(2N,\mathbb{R})$, i.e., $S$ is an element of a group isomorphic to the $Sp(2N,\mathbb{R})$, and $D$ is a diagonal matrix. $S$ and $D$ take the form
\[\label{def:S_and_D}
S=
\begin{bmatrix}
\A & \B \\
\ov{\B} & \ov{\A}
\end{bmatrix},\quad
D=\begin{bmatrix}
L & 0 \\
0 & L
\end{bmatrix},
\]
%S=\begin{bmatrix} \alpha & \beta \\ \overline{\beta} & \overline{\alpha} \end{bmatrix}, D=\begin{bmatrix} L & 0 \\ 0 & L \end{bmatrix}, L=\mathrm{\textbf{diag}}(\lambda_1,\dots,\lambda_N)
where $S$ additionally satisfies the relation $SKS^\dag=K$ with $K$ defined by Eq.~\eqref{def:commutation_relation}, and $L=\mathrm{\textbf{diag}}(\lambda_1,\dots,\lambda_N)$ is a diagonal matrix consisting of the so-called symplectic eigenvalues of a covariance matrix $\sigma$. This result will be used through this article.

The symplectic eigenvalues can be found by solving the usual eigenvalue problem for the matrix $K\sigma$. Eigenvalues of $K\sigma$ always appear in pairs. If $\lambda_i$ is an eigenvalue of $K\sigma$, then also $-\lambda_i$ is an eigenvalue of the same operator. The symplectic spectrum is then defined as a collection of the positive eigenvalues of $K\sigma$. In other words, $\lambda_i$ is a symplectic eigenvalue of $\sigma$ if and only if it is positive and $\pm\lambda_i$ are the eigenvalues of the operator $K\sigma$. Symplectic eigenvalues are always greater or equal to one and are related to the purity of the Gaussian state. The state is pure if and only if for all $i$, $\lambda_i=1$, and the larger the symplectic eigenvalues are, the more mixed the state is. Knowing this, we can say symplectic eigenvalues are analogous to the eigenvalues of the density matrix $\hat{\rho}$ in the density-matrix formalism. On the other hand, symplectic matrices $S$ usually represent some form of a squeezing or an entangling operation and are analogous to the unitary operators in the density matrix formalism. Given the special form in Eq.~\eqref{def:S_and_D} of the symplectic matrices and the relation $SKS^\dag=K$, one can easily prove~\cite{Mostafazadeh2004a} that the complex form of the symplectic matrices forms a subgroup of the more general pseudo-unitary group $U(N,N)=\{S\in GL(2N,\mathbb{C})|SKS^\dag=K\}$.

For more details about the complex and the real form of the covariance matrix see~\ref{app:real_covariance}, for a more detailed analysis of Gaussian states see~\cite{Adesso2014a,Weedbrook2012a}.

\section{Quantum estimation of two-mode Gaussian state}
In this section we derive an exact expression for the quantum Fisher information for any two-mode Gaussian state. There are numerous ways to compute this quantity, however, for the purpose of this section we adopt the definition via the Bures distance~\cite{Bures1969a}. The Bures distance is a measure of distinguishability between two quantum states $\rho_{1,2}$ and is defined through the Uhlmann fidelity~\cite{Uhlmann1976a} $\mathcal{F}({\rho}_{1},{\rho}_{2})\,:=\,\big(\tr\sqrt{\sqrt{{\rho}_{1}}\,{\rho}_{2}\,\sqrt{{\rho}_{1}}}\big)^{2}$ as $d_B^2=2\big(1-\sqrt{\mathcal{F}(\rho_1,\rho_2)}\big)$. The quantum Fisher information which measures how well we can distinguish two neighboring states $\rho_\epsilon$ and $\rho_{\epsilon+\de}$ is defined as a limit~\cite{Hayashi2006a}
\[\label{eqn:bures}
H(\epsilon)=8\lim_{\de\rightarrow0}\frac{1-\sqrt{\mathcal{F}(\rho_\epsilon,\rho_{\epsilon+\de})}}{\de^{2}}.
\]
The problem of finding the quantum Fisher information thus reduces to expanding the fidelity around the point $\epsilon$. As stated before, for Gaussian states the density matrix can be represented by a couple of the first and the second moments, $\rho_1\equiv(\boldsymbol{d}_1,\sigma_1)$, $\rho_2\equiv(\boldsymbol{d}_2,\sigma_2)$. In the case of a two-mode Gaussian state the fidelity can be written as~\cite{Marian2012a}
\[\label{eq:fidelity_basic_formula}
\mathcal{F}(\rho_1,\rho_2)=\frac{4e^{-\delta\boldsymbol{d}^{\dagger}\left(\sigma_{1}
+\sigma_2\right)^{-1}\delta\boldsymbol{d}}}{\left(\sqrt{\Gamma}+\sqrt{\Lambda}\right)-\sqrt{\left(\sqrt{\Gamma}+\sqrt{\Lambda}\right)^{2}-\Delta}},
\]
where $\delta\boldsymbol{d}=\boldsymbol{d}_1-\boldsymbol{d}_2$ is the relative displacement and $\Delta,\Gamma,\Lambda$ denotes three determinants defined as
\begin{subequations}\label{eqs:GDL_unpolished}
\begin{align}
\Delta &=\det{\sigma_{1}+\sigma_{2}},\\
\Gamma &=\det{I+K\sigma_{1}K\sigma_{2}},\\
\Lambda &=\det{\sigma_{1}+K}\det{\sigma_{2}+K},
\end{align}
\end{subequations}
with $K=I\oplus{-I}$ already introduced in the previous section. We note that our definitions differs from~\cite{Marian2012a} by the factor of $2$, which results in the factor of $4$ in Eq.~\eqref{eq:fidelity_basic_formula}.

Let us denote an expansion of an arbitrary matrix around point $\epsilon$ up to second order in $\de$ as
\[\label{eqn:matrix_not}
M=M_0+M_1\de+M_2\de^2+O(\de^3),
\]
where $M=M(\epsilon+\de)$, $M_0=M(\epsilon)$, $M_1=\dot{M}(\epsilon)$, $M_2=\frac{1}{2}\ddot{M}(\epsilon)$, where \emph{dot} denotes the derivative with respect to $\epsilon$. Using this notation and the definition~\eqref{eqn:bures} we can write $H(\epsilon)=-4\mathcal{F}_2(\epsilon)$, where $\mathcal{F}$ from now on denotes the fidelity between the two close states $\rho_\epsilon$ and $\rho_{\epsilon+\de}$. The problem is that if we try to use Eqs.~\eqref{eqs:GDL_unpolished} to expand Eq.~\eqref{eqn:bures} directly to find the expression of $\mathcal{F}_2$, we arrive at a complicated expression that depends on the second derivatives of $\sigma$. However, expressions of the quantum Fisher information in~\cite{Paris2009a,Monras2013a} do not depend on second derivatives. This is true most of the time but there are places where the second derivatives appear. This fact is usually ignored in the literature and we will explain those scenarios later. First, to find an expression that depends only on first derivatives we use Williamson's theorem $\sigma=SDS^\dag$ to rewrite Eqs.~\eqref{eqs:GDL_unpolished} as
\begin{subequations}\label{eqs:GDL_polished}
\begin{align}
\Delta &= \det{D_0+PDP^\dagger},\\
\Gamma &= \det{I+D_0PDP^\dagger},\\
\Lambda &= \det{D_0+K}\det{D+K},
\end{align}
\end{subequations}
where the matrix $P$ is defined as $P=S_0^{-1}S$. Following Eq.~\eqref{eqn:matrix_not}, $P$ has expansion $P=I+P_1\de+P_2\de^2+O(\de^3)$, where $P_1=S_0^{-1}S_1=S(\epsilon)^{-1}\dot{S}(\epsilon)$ and $P_2=S_0^{-1}S_2$. A useful property of $P$ is that it is symplectic, i.e., $PKP^\dag=K$, giving us conditions on the first and second derivatives,
\begin{subequations}\label{eq:P_symplectic}
\begin{align}
P_1^\dag&=-KP_1K\label{id:P_1_symplectic},\\
P_2^\dag&=-KP_2K+KP_1^2K.
\end{align}
\end{subequations}
Using the expansion of the determinant,
\begin{multline}
\det{M_0+M_1\de+M_2\de^2+O(\de^3)}=\det{M_0}\Big(1+\left(\tr M_0^{-1}M_1\right)\de\\
+\frac{1}{2}\!\Big(\!2\tr M_0^{-1}M_2+(\tr M_0^{-1}M_1)^2\!\!-\tr [(M_0^{-1}M_1)^2]\!\Big)\de^2\!\Big)+O(\de^3),
\end{multline}
which holds for an invertible matrix $M_0$, equations~\eqref{eq:P_symplectic}, and the cyclic property of the trace we remove the dependence on second derivatives $P_2$ and obtain
\begin{subequations}\label{eqs:DeltaGammaLambda}
\begin{align}
\Delta&=\det{D_0}\Big(16+8\tr[D_0^{-1}D_1]\de+\Big(4\tr[P_1^2]\nonumber\\
&-4\tr[D_0^{-1}KP_1D_0KP_1]+8\tr[D_0^{-1}D_2]\\
&+2(\tr [D_0^{-1}D_1])^2-2\tr [(D_0^{-1}D_1)^2]\Big)\de^2\Big)+O(\de^3),\nonumber\\
\Gamma&=\det{C}\Big(1+\tr [C^{-1}D_0D_1]\de+\Big(\tr[C^{-1}P_1^2]\nonumber\\
&-\tr[(C^{-1}P_1)^2]-\tr[(C^{-1}D_0KP_1)^2]\\
&+\tr[C^{-1}D_0D_2]+\frac{1}{2}(\tr [C^{-1}D_0D_1])^2\nonumber\\
&-\frac{1}{2}\tr [(C^{-1}D_0D_1)^2]\Big)\de^2\Big)+O(\de^3),\nonumber\\
\Lambda&=\det{E}^2\Big(1+\tr[E^{-1}D_1]\de+\Big(\tr [E^{-1}D_2]\\
&+\frac{1}{2}(\tr[ E^{-1}D_1])^2-\frac{1}{2}\tr [(E^{-1}D_1)^2]\Big)\de^2\Big)+O(\de^3),\nonumber
\end{align}
\end{subequations}
where we have denoted $C=I+D_0^2$ and $E=D_0+K$. In the above we assumed that there exist a Taylor expansion around point $\epsilon$ in this particular form, which is not true for $\Lambda$ in the case where at least one of the symplectic eigenvalues of $\sigma$ is equal to one. We address this subtle issue in~\ref{app:pure_regularization}. Assuming all symplectic eigenvalues are larger than one, we insert expressions \eqref{eqs:DeltaGammaLambda} into the Uhlmann fidelity \eqref{eq:fidelity_basic_formula}. We derive that the zeroth order sums to 1, the first order vanishes, and the second order provides the quantum Fisher information
\[
H(\epsilon)=\frac{\Delta_2-8((\sqrt{\Gamma})_2+(\sqrt{\Lambda})_2)}{2(\sqrt{\Gamma_0}+\sqrt{\Lambda_0}-4)}+2\bd_1^\dag\sigma_0^{-1}\bd_1.\\
\]
The denominator in the above expression actually has a compact expression in terms of either a) the determinant of the state or b) the symplectic eigenvalues, $\sqrt{\Gamma_0}+\sqrt{\Lambda_0}-4=2(\det{\sigma}-1)=2(\det{D_0}-1)$, which helps with computations considerably. To derive an expression which depends only on the first derivatives, we insert Eqs.~\eqref{eqs:DeltaGammaLambda} into the above formula. The terms proportional to $D_2$ vanish giving us an alternative expression for the quantum Fisher information,
\[\label{GeneralQFISD}
\begin{split}
H&(\epsilon)=\frac{1}{\det{D_0}-1}\bigg(\det{D_0}\Big(\tr[P_1^2]-\tr[D_0^{-1}KP_1D_0KP_1]\Big)\\
&\!+\!\sqrt{\det{C}}\Big(\tr[(C^{-1}P_1)^2]\!+\!\tr[(C^{-1}D_0KP_1)^2]\!-\!\tr[C^{-1}P_1^2]\Big)\!\bigg)\\
&+\frac{1}{2}\tr\big[(D_0+K)^{-1}D_0^{-1}D_1^2\big]+2\bd_1^\dag\sigma_0^{-1}\bd_1,
\end{split}
\]
where ${P_1=S_0^{-1}S_1=S(\epsilon)^{-1}\dot{S}(\epsilon)}$, $D_0=D(\epsilon)$, $D_1=\dot{D}(\epsilon)$, $C=I+D_0^2$, and $\bd_1=\dot{\bd}(\epsilon)$. The above formula is useful when we know the initial Williamson decomposition. For example when we are trying to estimate squeezing in the case of two-mode squeezed thermal state. However, in general finding the Williamson decomposition is not an easy task. That is why we find an alternative expression only in terms of the covariance matrix $\sigma$, displacement $\bd$ and symplectic eigenvalues. For convenience we use the \emph{dot}-notation, where \emph{dot} denotes the derivative with respect to $\epsilon$. For convenience we also denote $A:=K\sigma(\epsilon)$. The quantum Fisher information for a two-mode Gaussian state is given by
\begin{widetext}
\[\label{GeneralQFI}
H(\epsilon)=\frac{1}{2(\det{A}-1)}\Bigg(\det{A}\tr\Big[\big(A^{-1}\dot{A}\big)^2\Big]+\sqrt{\det{I+A^2}}\tr\Big[\big((I+A^2)^{-1}\dot{A}\big)^2\Big]
+4\big(\lambda_1^2-\lambda_2^2\big
)\bigg(-\frac{\dot{\lambda_1}^2}{\lambda_1^4-1}
+\frac{\dot{\lambda_2}^2}{\lambda_2^4-1}\bigg)\Bigg)+2\dot{\bd}^\dag\sigma^{-1}\dot{\bd},
\]
\end{widetext}
where the symplectic eigenvalues of $\sigma$ can be calculated as \[\lambda_{1,2}=\frac{1}{2}\sqrt{\tr[A^2]\pm\sqrt{(\tr[A^2])^2-16\det{A}}},\] and $\det{\cdot}$ denotes the determinant. Using the Williamson theorem one can prove that the above formula reduces to Eq.~\eqref{GeneralQFISD}. Also, strictly speaking, the above formula is defined only for covariance matrices with both symplectic eigenvalues larger than one. However, we can use a regularization procedure which allows us to use this formula in any case. This consists of multiplying the original covariance matrix $\sigma$ by a mixedness parameter $\nu>1$, using the formula~\eqref{GeneralQFI} to calculate the quantum Fisher information for the state $\nu\sigma$, performing the limit $\nu\longrightarrow1$, and adding the second derivative $\ddot{\lambda_i}(\epsilon)$ for every symplectic eigenvalue ${\lambda_i}(\epsilon)$ which is equal to one. We need to add these second derivatives because by performing the limit $\nu\rightarrow 1$ we set the problematic terms $\frac{\dot{\lambda_i}^2}{\lambda_i^4-1}$ of Eq.~\eqref{GeneralQFI} to zero. However, in cases when the symplectic eigenvalue $\lambda_i$ is equal to one, such terms have a non-zero contribution which needs to be accounted for. Altogether, we have
\[\label{eq:regularization_proc}
H(\epsilon)=\lim_{\nu\rightarrow1}H\big(\nu\sigma(\epsilon)\big)+\!\!\!\!\!\!\sum_{i:\lambda_i(\epsilon)=1}\!\!\!\!\!\!\ddot{\lambda}_i(\epsilon).
\]
More details about the procedure can be found in~\ref{app:pure_regularization}.

Note that in the case where the symplectic eigenvalues do not change with a small variation in $\epsilon$, i.e. $\dot{\lambda_1}=\dot{\lambda_2}=0$, or $\ddot{\lambda_i}=0$ for ${\lambda_i}=1$, the term depending on the symplectic eigenvalues vanishes. This includes the case where purity does not change or where the parameter of interest $\epsilon$ was encoded into the initial state by a symplectic transformation. Although all the computations were performed in the complex representation of the covariance matrix, we can easily transform the result to the real representation. This is done by substituting $A=i\Omega\sigma_R$, $\dot{\bd}^\dag\sigma^{-1}\dot{\bd}\longrightarrow \dot{\bd}_R^T\sigma_R^{-1}\dot{\bd}_R$, where $i\Omega$ is the real form equivalent to matrix $K$, $\sigma_R$ the real form covariance matrix and $\bd_R$ the real form displacement. For more details see~\ref{app:real_covariance}.

\section{Multi-mode parameter estimation}
In the previous section we derived an exact expression for two-mode Gaussian states, however, in recent work~\cite{Monras2013a} a general formula for the quantum Fisher information was derived as a limit of a particular infinite series. Here we simplify the expression for the infinite series. We also find a bound on the remainder of the series when we sum only a finite number of terms. We then go on to simplify an already known formula for pure states and derive an exact expression for the multi-mode quantum Fisher information for the cases when the Williamson decomposition is known.

According to~\cite{Monras2013a}, the quantum Fisher information for a general multi-mode case which has all symplectic eigenvalues larger than one can be calculated as
\[\label{eq:Monras_QFI}
H(\epsilon)=\frac{1}{2}\tr\big[\dot{\sigma}Y\big]+2\dot{\bd}^\dag\sigma^{-1}\dot{\bd},
\]
where $Y$ is a solution to the so-called Stein equation~\cite{Bhatia2006a}
\[
\dot{\sigma}=\sigma Y\sigma-KYK.
\]
If all symplectic eigenvalues of $\sigma$ are larger than one, the solution is unique and can be written as an infinite series
\[\label{eq:solution}
Y=-\sum_{n=0}^\infty (K\sigma)^{-n}\dot{(\sigma^{-1})}(\sigma K)^{-n}.
\]
Inserting \eqref{eq:solution} into \eqref{eq:Monras_QFI}, using $\dot{(\sigma^{-1})}=-\sigma^{-1}\dot{\sigma}\sigma^{-1}$, and the properties of matrix $K$ we find an elegant expression for the multi-mode quantum Fisher information,
\[\label{eq:Monras_polished}
H(\epsilon)=\frac{1}{2}\sum_{n=1}^M\tr\big[(A^{-n}\dot{A})^2]+R_M+2\dot{\bd}^\dag\sigma^{-1}\dot{\bd},
\]
where we use the notation $A:=K\sigma$ or $A:=i\Omega\sigma_R$ for the real form of the covariance matrix and $R_M$ is the remainder of the series $M\longrightarrow\infty$. As we prove in~\ref{app:Remainder}, the remainder is bounded $|R_M|\leq\frac{\mathrm{tr}[(A\dot{A})^2]}{2\lambda_{\mathrm{min}}^{2M+2}(\lambda_{\mathrm{min}}^2-1)}$ with $\lambda_{\mathrm{min}}:=\min_{i}\{\lambda_i\}$ being the smallest symplectic eigenvalue, i.e., the smallest positive eigenvalue of matrix $A$. This means that for $\lambda_{\mathrm{min}}>1$ we have $\lim_{M\longrightarrow\infty}R_M=0$, the series converges, and we can write
\[\label{eq:Monras_polished_infty}
H(\epsilon)=\frac{1}{2}\sum_{n=1}^\infty\tr\big[(A^{-n}\dot{A})^2]+2\dot{\bd}^\dag\sigma^{-1}\dot{\bd}.
\]
To calculate the quantum Fisher information for the states which have some eigenvalues equal to one we can, once again, use the regularization procedure~\eqref{eq:regularization_proc}. A small example, which has been already shown in~\cite{Monras2013a}, is to consider the class of iso-thermal states given by $A^2=\nu^2I$, $\nu>1$, which is equivalent to $D(\epsilon)=\nu I$. For such states we can easily evaluate the infinite sum \eqref{eq:Monras_polished_infty} and derive
\[\label{eq:nu_pure}
H(\epsilon)=\frac{\nu^{2}}{2(1+\nu^{2})}\mathrm{tr}\big[(A^{-1}\dot{A})^2\big]+2\dot{\bd}^\dag\sigma^{-1}\dot{\bd},
\]
which for $\nu=1$ gives a formula for pure states also derived in~\cite{Pinel2012a}. As noted in~\cite{Jiang2014a}, this formula can be further simplified. Differentiating $A^2=\nu^2$ and multiplying each side by $A^{-1}$ we obtain an anti-commutation relation $A^{-1}\dot{A}=-\dot{A}A^{-1}$ which together with $A^{-2}=\nu^{-2}I$ gives
\[
H(\epsilon)=-\frac{1}{2(1+\nu^{2})}\mathrm{tr}\big[\dot{A}^2\big]+2\dot{\bd}^\dag\sigma^{-1}\dot{\bd},
\]
meaning that for iso-thermal states with fixed displacement we do not need to invert the covariance matrix anymore.

Using $\tr[(A^{-n}\dot{A})^2]=2\tr[(D_0^{-n+1}K^{-n+1}P_1)^2]-2\tr[D_0^{-n+2}K^{-n}P_1D_0^{-n}K^{-n}P_1]+\tr[D_0^{-2n}D_1^2]$ we can rewrite formula \eqref{eq:Monras_polished_infty} for the quantum Fisher information in terms of the elements of the symplectic matrices and eigenvalues,
\[\label{eq:multimode_P1}
\begin{split}
H(\epsilon)&=\tr\big[P_1^2\big]-\sum_{n=1}^\infty\tr\Big[\big(D_0^2-I\big)\big(D_0^{-n}K^{-n}P_1\big)^2\Big]\\
&+\frac{1}{2}\tr\big[(D_0+K)^{-1}D_0^{-1}D_1^2\big]+2\bd_1^\dag\sigma_0^{-1}\bd_1,
\end{split}
\]
where we have used the notation from Eq.~\eqref{GeneralQFISD}. We can see that the diagonal part coincides with the two-mode case indicating the validity of the general multi-mode formula. However, in contrast to Eq.~\eqref{eq:Monras_polished_infty} the infinite series here converges also when symplectic eigenvalues are equal to one. Nevertheless, this does not mean that it is valid to use this formula for such states. To be more specific, plainly inserting $D_0=I$ does \emph{not} give the correct formula for pure states --- this still needs to be obtained by the regularization procedure. The reason why it does not give the correct result is that in general limits $\lim_{M\rightarrow\infty}$ and $\lim_{\nu\rightarrow1}$ do not commute. On the other hand, the infinite sum in this formula leads to a geometric series that can be evaluated, as shown in~\ref{app:exact_multimode}. This allows us to derive a much more elegant, entirely general and exact formula for the quantum Fisher information in terms of elements of the Williamson decomposition $\sigma=SDS^\dag$ for multi-mode Gausian states. Using the definitions~\eqref{def:S_and_D} and \eqref{id:P_1_symplectic}, we find that matrix $P_1$
has an elegant structure
\[\label{def:S_and_P}
P_1(\epsilon)=
\begin{bmatrix}
R & Q \\
\ov{Q} & \ov{R}
\end{bmatrix},
\]
where $R=\A^\dag\dot{\A}-\overline{\B^\dag\dot{\B}}$ is a skew-Hermitian and $Q=\A^\dag\dot{\B}-\overline{\B^\dag\dot{\A}}$ a (complex) symmetric matrix. Matrix $P_1$ is actually an element of the Lie algebra associated with the complex form of the real symplectic group. If the diagonalizing symplectic matrix forms a one-parameter group $S=e^{X\epsilon}$, where $X$ is an element of the algebra independent of $\epsilon$, then $P_1=X$. Inserting Eq.~\eqref{def:S_and_P} into Eq.~\eqref{eq:multimode_P1} and evaluating the infinite sum we derive the quantum Fisher information for the $N$-mode Gaussian state
\[\label{eq:exact_multimode}
\begin{split}
H(\epsilon)&=\sum_{i,j=1}^N\frac{(\lambda_i-\lambda_j)^2}{\lambda_i\lambda_j-1}\norm{R_{ij}}^2+\frac{(\lambda_i+\lambda_j)^2}{\lambda_i\lambda_j+1}\norm{Q_{ij}}^2\\
&+\sum_{i=1}^N\frac{\dot{\lambda_i}^2}{\lambda_i^2-1}+2\dot{\bd}^\dag\sigma^{-1}\dot{\bd},
\end{split}
\]
%H(\epsilon)=\sum_{i,j=1}^N\frac{(\lambda_i-\lambda_j)^2}{\lambda_i\lambda_j-1}|R_{ij}|^2+\frac{(\lambda_i+\lambda_j)^2}{\lambda_i\lambda_j+1}|Q_{ij}|^2+\sum_{i=1}^N\frac{\dot{\lambda_i}^2}{\lambda_i^2-1}+2\dot{d}^\dag\sigma^{-1}\dot{d}
where $\lambda_i$ are the symplectic eigenvalues. Strictly speaking, the formula is not defined for the symplectic eigenvalues equal to one. Under the assumption of differentiability of the covariance matrix, for $\lambda_i(\epsilon)=\lambda_j(\epsilon)=1$ both $\frac{(\lambda_i-\lambda_j)^2}{\lambda_i\lambda_j-1}(\epsilon)$ and $\frac{\dot{\lambda_i}^2}{\lambda_i^2-1}(\epsilon)$ are expressions of type $\frac{0}{0}$. Nevertheless, we can define these problematic points in a way which makes the quantum Fisher information a continuous function, which also corresponds to the knowledge gained from a study of the states close to the points of purity described in~\ref{app:pure_regularization}. For $\lambda_i(\epsilon)=1$ we define $\frac{\dot{\lambda_i}^2}{\lambda_i^2-1}(\epsilon):=\ddot{\lambda_i}(\epsilon)$, and for $\lambda_i(\epsilon)=\lambda_j(\epsilon)=1$ we define $\frac{(\lambda_i-\lambda_j)^2}{\lambda_i\lambda_j-1}(\epsilon):=0$.

Assuming all symplectic eigenvalues are larger than one, we can define the Hermitian matrix $\widetilde{R}_{ij}:=\frac{\lambda_i-\lambda_j}{\sqrt{\lambda_i\lambda_j-1}}R_{ij}$ and the symmetric matrix $\widetilde{Q}_{ij}:=\frac{\lambda_i+\lambda_j}{\sqrt{\lambda_i\lambda_j+1}}Q_{ij}$ which allows us to write the quantum Fisher information in a compact form,
\[\label{eq:exact_multimode_compact}
H(\epsilon)=\tr\big[\widetilde{R}\widetilde{R}^\dag+\widetilde{Q}\widetilde{Q}^\dag\big]+\tr\big[(L^2-I)^{-1}\dot{L}^2\big]+2\dot{\bd}^\dag\sigma^{-1}\dot{\bd}.
\]
Now we can easily interpret each term. The first part corresponds to the change in relative orientation and squeezing, the second to the change in purity and the third to the change in displacement. Added together, they all contribute to the quantum Fisher information and increase the precision in the estimation of a parameter $\epsilon$.

Note that formulas \eqref{eq:exact_multimode}, \eqref{eq:exact_multimode_compact} respectively, are actually multi-mode generalizations of Eq.~\eqref{GeneralQFISD}. The form \eqref{GeneralQFISD} takes when we rewrite it in terms of submatrices given by Eq.~\eqref{def:S_and_P} is exactly the same as \eqref{eq:exact_multimode} for $N=2$. The same holds for the one-mode formula derived in~\cite{Pinel2013b} and $N=1$, partially validating our general result. For the derivation of the formula see~\ref{app:exact_multimode}, for the details why we choose that particular definition of problematic points see~\ref{app:pure_regularization}, and for the quantum Fisher information in the real form formalism see~\ref{app:real_covariance}.

\section{Example}

\begin{figure}[t!]
\centering
\includegraphics[width=1\linewidth]{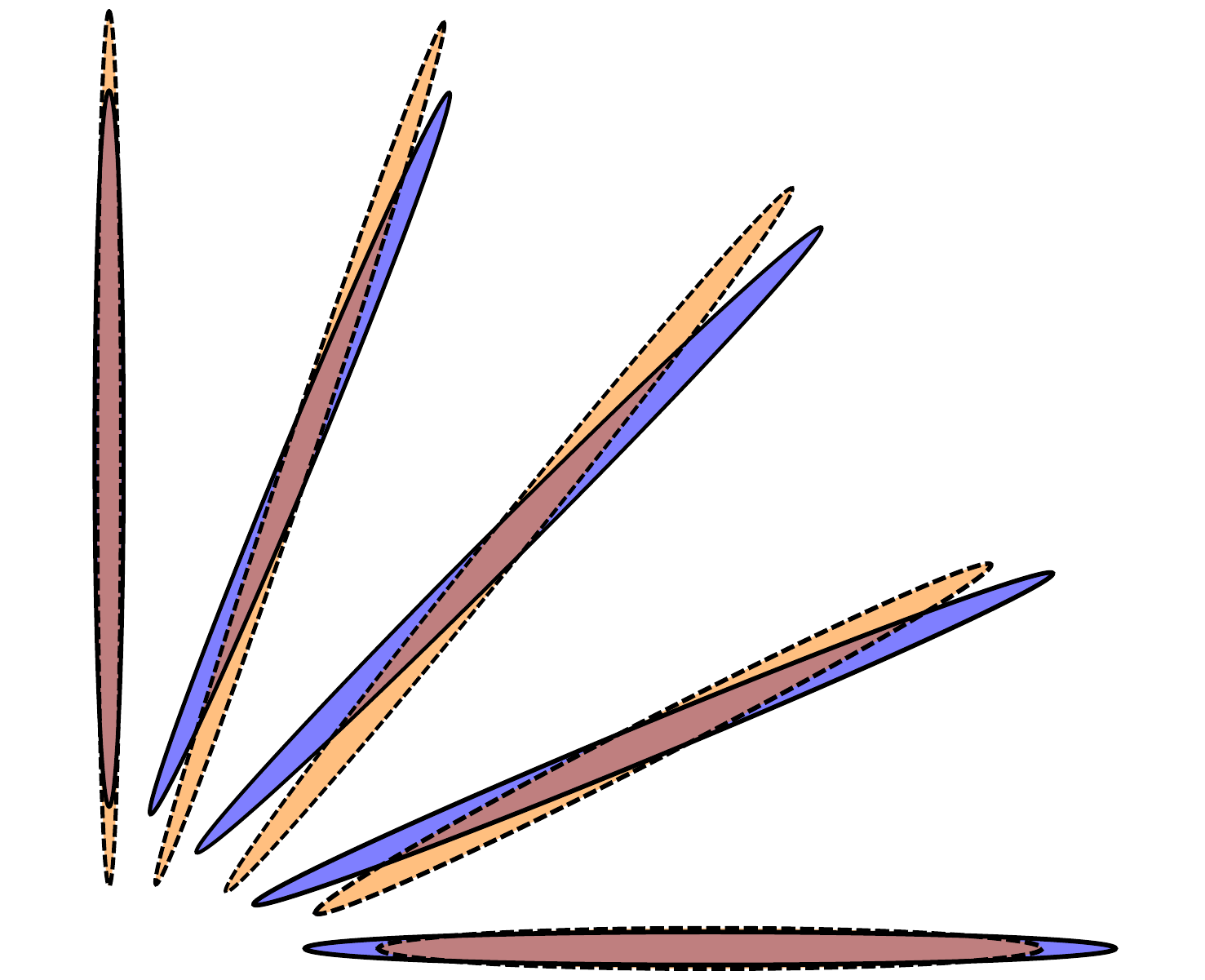}
\caption{Visualization of the distinguishability of the covariance matrices with different initial rotations. The initial squeezing was set to $r=0.8$, the initial displacement $\tilde{d_0}=0$, and the final squeezing $\epsilon=0$~(blue with full line) or $\epsilon=0.1$~(orange with dashed line). The initial rotation from left to right $\theta=0,\frac{\pi}{8},\frac{\pi}{4},\frac{3\pi}{8},\frac{\pi}{2}$. Covariance matrices with $\theta=\frac{\pi}{4}$ can be easily distinguished allowing for a better estimation of the parameter $\epsilon$.}\label{fig:three_cov}
\end{figure}

To illustrate the use of the general formula~\eqref{eq:exact_multimode} we derive the precision with which we can estimate the squeezing parameter $\epsilon$ of a one-mode squeezing channel. Similar problems have already been studied. For example in~\cite{Milburn1994a} the vacuum and coherent probe states were considered, in~\cite{Chiribella2006a} it was the coherent states and the displaced squeezed vacuum states, which were squeezed in either position or momenta direction. Here we are generalizing the precision bounds in those articles. We also note that a similar work has been done in~\cite{Gaiba2009a} using the density matrix formalism for a different type of a one-mode squeezing channel $\tilde{S}_r=\exp(-i\frac{r}{2}(\hat{a}^2+\hat{a}^{\dag2}))$. Here, however, we use an entirely general single-mode Gaussian state as an input state.

We choose a one-mode squeezed rotated displaced thermal state as an input state, which is the most general single-mode state~\cite{Weedbrook2012a}, with the initial squeezing parameter $r$, angle of rotation $\theta$, and the initial displacement $\tilde{d_0}=\norm{\tilde{d_0}}e^{i\phi}$. We feed this state into a one-mode squeezing channel, which encodes the unknown parameter $\epsilon$ we are trying to estimate, leaving us with the final state,
\[
\rho_\epsilon=S_\epsilon D_{\tilde{d_0}}R_\theta S_r\rho_{th}S_r^\dag R_\theta^\dag D_{\tilde{d_0}}^\dag S_\epsilon^\dag,
\]
where $S_r=\exp(\frac{r}{2}(\hat{a}^2-\hat{a}^{\dag2}))$ is the one-mode squeezing operator,
$R_\theta=\exp(-i\theta\hat{a}^{\dag}\hat{a})$ the rotation operator, and $D_{\tilde{d_0}}=\exp({\tilde{d_0}}\hat{a}^\dag-\ov{{\tilde{d_0}}}\hat{a})$ the displacement operator. As we show in~\ref{app:example}, the quantum Fisher information is
\[\label{eq:QFI_squeeze_gen}
\begin{split}
H(\epsilon)\!&=\!\frac{4\lambda_1^2}{\lambda_1^2+1}\big(\cosh^4r+\sinh^4r-2\cos(4\theta)\cosh^2r\sinh^2r\big)\\
&+\frac{4\norm{\tilde{d_0}}^2}{\lambda_1}\big(\cosh(2r)+\cos(2(\theta-\phi))\sinh(2r)\big).
\end{split}
\]
Now, let us have a few notes on the derived formula. The result does not depend on the parameter we want to estimate, which must be true for every encoding channel which forms a one-parameter unitary group $U(\epsilon)=e^{-i\hat{K}\epsilon}$, where $\hat{K}$ is a Hermitian operator. Also, as pointed out in~\cite{Aspachs2008a}, because the symplectic eigenvalue $\lambda_1\geq1$ is proportional to temperature, we can immediately see that although the thermality slightly enhances the estimation through squeezing (with the maximal enhancement by the factor of 2), it reduces the estimation from the displacement. More importantly, however, by choosing an appropriate rotation of an input state we can significantly increase the estimation precision. Without loss of generality we assume $r\geq0$. The maximal value of~\eqref{eq:QFI_squeeze_gen} is achieved when we choose to rotate for example by the value $\theta=\frac{\pi}{4}$, and displace in the direction $\phi=\frac{\pi}{4}$,
\[\label{eq:optimal_QFI_squeeze}
H_{\mathrm{max}}(\epsilon)=\frac{4\lambda_1^2}{\lambda_1^2+1}\cosh^2(2r)+\frac{4}{\lambda_1}\norm{\tilde{d_0}}^2e^{2r},
\]
which signifies an exponential increase in the precision of estimation the unknown parameter $\epsilon$ when considering an initially squeezed and rotated state. To demonstrate, we derive a formula for a squeezing needed to enhance by $k$ orders of magnitude when assuming zero initial displacement, $r=\mathrm{arcsinh}\sqrt{\frac{10^{\frac{k}{2}}-1}{2}}$, which for larger $k$ (or $r\gtrsim1$) behaves as $r\approx0.35+0.58k$. The current state-of-the-art~\cite{Eberle2010a} achieves the squeezing around $r=1.46$, which could hypothetically account for an improvement by a factor of $80$ as compared to the case where the initial squeezing is zero. As we show in figure~\ref{fig:three_cov}, the reason why the amount of distinguishability of the two close states rises is because the final squeezing forces the covariance matrix to turn. Also, note that although the initial squeezing leads to an exponential increase, the initial displacement contributes only quadratically. To conclude, the optimal Gaussian state for the estimation of the parameter of a one-mode squeezing channel is a thermal state infinitely squeezed in the angle of $\theta=\frac{\pi}{4}$ from the direction of the squeezing channel we want to estimate, and infinitely displaced in the direction in which the squeezed state is stretched,  $\phi=\frac{\pi}{4}$. For a fixed amount of squeezing and displacement, the optimal temperature which maximizes the quantum Fisher information is given by a solution of
$\frac{\lambda_1^3}{(\lambda_1^2+1)^2}=\frac{\norm{\tilde{d_0}}^2e^{2r}}{2\cosh^2(2r)}$. Also, because $\lambda_1=1+2n_{th}$, where $n_{th}=\tr[\rho_{th}\hat{a}^\dag\hat{a}]$ denotes the mean number of thermal bosons, and the mean total number of bosons in displaced squeezed states is given by
\[
n=n_{\tilde{d_0}}+n_{th}+(1+2n_{th})\sinh^2r,
\]
where we have denoted $n_{\tilde{d_0}}:=\norm{\tilde{d_0}}^2$ the mean number of bosons coming from displacement, we can rewrite the maximal quantum Fisher information~\eqref{eq:optimal_QFI_squeeze} as
\[
\begin{split}
&H_{\mathrm{max}}(\epsilon)=\frac{2(1+2n-2n_{\tilde{d_0}})^2}{1+2n_{th}(1+n_{th})}\\
&\ +\frac{4n_{\tilde{d_0}}\left(1+2n-2n_{\tilde{d_0}}+2\sqrt{n\!-\!n_{\tilde{d_0}}\!-\!n_{th}}\sqrt{1\!+\!n\!-\!n_{\tilde{d_0}}\!+\!n_{th}}\right)}{(1+2n_{th})^2}.
\end{split}
\]
For a fixed total number of bosons this function achieves the maximum at $n_{th}=n_{\tilde{d_0}}=0$. This means that for a finite amount of available energy, the best protocol is to invest it all into squeezing, which is the same result as in~\cite{Aspachs2008a,Gaiba2009a}. The optimal probe state is then the $\frac{\pi}{4}$-rotated squeezed vacuum state with the quantum Fisher information $H(\epsilon)=2(1+2n)^2$, which clearly indicates Heisenberg scaling.

\section{Conclusion}
We have derived an exact formula for the quantum Fisher information of an arbitrary two-mode Gaussian state. This has been done using the definition of the infinitesimal Bures distance, the Williamson decomposition of positive-definite matrices and the properties of the real symplectic group. Although the formula is not directly applicable for states with pure modes, we introduced a regularization procedure which allows us to overcome this problem. Then, using a different approach, we simplified an already known formula for the multi-mode quantum Fisher information in terms of an infinite series. We also estimated the remainder of the series, allowing for an effective numerical calculation. Using the previous result, we showed that for the cases when the Williamson decomposition of the covariance matrix is known, the quantum Fisher information for multi-mode Gaussian state can be computed exactly. The general multi-mode formula is equivalent to the known results for one-mode Gaussian state when setting $N=1$ and to the previously mentioned two-mode Gaussian states when $N=2$. However, we note that using the requirement of the continuity of the quantum Fisher information and studying the case of the pure states gave us a different definition for the problematic points than is mentioned in~\cite{Pinel2013b}. Finally we applied our newly gained formula to study the case of the estimation a squeezing parameter of a one-mode squeezing channel. We showed that a strategy of squeezing and rotating the input state can significantly improve the precision in estimation.

We believe the main achievement of this article is in the usefulness of the derived formulae. It allows for the study of the optimal input states of Gaussian nature~\cite{Invernizzi2011a}, it helps predict the ultimate sensitivity of a physical detector's particular implementation~\cite{Caron1995a,Abbott2004a}, or to analyse the effects of temperature on the current gravitational wave detector proposal~\cite{Sabin2014a}. It gives a limit in the estimation of time~\cite{Lindkvist2014a} or temperature~\cite{Correa2014a}. Also, since certain objects called the Bogoliubov transformations are isomorphic to the symplectic transformations, the natural application lies wherever these transformations appear. This is for example for quantum field theory in curved spacetime~\cite{Birrell1984a,Friis2015a} but also Bose-Einstein condensates~\cite{Utsunomiya2008a} or scattering problems~\cite{Boonserm2009a}.

\emph{Acknowledgement} Thanks to Carlos Sab\'{i}n, M\u{a}d\u{a}lin Gu\c{t}\u{a}, Alex Monras, David Edward Bruschi, Jan Kohlrus, and Katarzyna Macieszczak for fruitful discussions. A.~R.~L. thanks the EPSRC Doctoral prize. I.~F. acknowledges support from EPSRC INSPIRE:~EP/M003019/1.

\appendix
\section{The complex and the real form of the covariance matrix}\label{app:real_covariance}
In this section we describe the structure of covariance matrices and displacement. We introduce the real form covariance matrix formalism and find how covariance matrices in the real form transform into its complex form.

The complex form of the covariance matrix and displacement was defined by equations \eqref{def:commutation_relation} and \eqref{def:covariance_matrix}. From the definition we can observe the following structure of the first and second moments:
\[\label{def:first_and_second_moments}
\bd=
\begin{bmatrix}
\boldsymbol{\tilde{d}} \\ \overline{\boldsymbol{\tilde{d}}}
\end{bmatrix},\quad
\sigma\,=\,\begin{bmatrix}
X & Y \\
\overline{Y} & \overline{X}
\end{bmatrix}
\]
and $\sigma^\dag=\sigma$, i.e., $X^\dag=X$ and $Y^T=Y$. Using the Williamson theorem~\cite{Williamson1936a,Arvind1995a,Simon1998a} we can write $\sigma=SDS^\dag$, where $SKS^\dag=K$. Although the symplectic matrices $S$ are not necessarily Hermitian, they follow the same structure as $\sigma$ which is expressed by Eq.~\eqref{def:S_and_D}.

Construction of the real form of the covariance matrix is analogous to the complex form described in the introduction. It is usually defined with respect to the collection of quadrature operators $\boldsymbol{\hat{Q}}:=\boldsymbol{\hat{x}}\oplus\boldsymbol{\hat{p}}\,=\,\{\hat{x}_{1},\hat{x}_{2},\ldots,\hat{p}_{1},\hat{p}_{2},\ldots\}$. The canonical commutation relations of these operators can be conveniently expressed as
\begin{equation}
[\hat{{Q}}_{i},\hat{{Q}}_{j}]\,=+i\,\Omega_{ij}\,\mathrm{id}\quad\Rightarrow\quad \Omega=
\begin{bmatrix}
0 & I \\
-I & 0
\end{bmatrix}.
\end{equation}
Properties of $\Omega$ are $-\Omega^2=I$ and $\Omega^T=-\Omega$, in contrast to the complex form version $K$. In the real form, the definitions of the first and second moments are
\begin{subequations}
\begin{align}
\bd_{R}&=\mathrm{tr}\big[\hat{\rho}\boldsymbol{\hat{Q}}\big]=\begin{bmatrix}\boldsymbol{x} \\ \boldsymbol{p}\end{bmatrix},\\
\sigma_{R}&=\mathrm{tr}\big[\hat{\rho}\{\boldsymbol{\hat{Q}},\boldsymbol{\hat{Q}}\}\big]=\begin{bmatrix}X_{R} & Y_{R} \\ Y^{T}_{R} & Z_{R} \end{bmatrix}.
\end{align}
\end{subequations}
The real covariance matrix is symmetric, i.e. $X_R=X_R^T$ and $Z_R=Z_R^T$. The corresponding real symplectic matrices are given by $\sigma_R=S_RD_RS_R^T$, where $S\Omega S^T=\Omega$, which is a defining relation of the real symplectic group $Sp(2N,\mathbb{R})$.

Since the change between real and complex form of the covariance matrix is a simple basis transformation, $\boldsymbol{\hat{Q}}\rightarrow\boldsymbol{\hat{A}}$, we can relate these two using the unitary matrix $U$,
\[
\boldsymbol{\hat{A}}=U\boldsymbol{\hat{Q}},\ \ U=\frac{1}{\sqrt{2}}\,\begin{bmatrix}I & +iI \\ I & -iI\end{bmatrix}.
\]
The resulting transformation between real and complex covariance matrices and displacement are
\[
\bd=U\bd_{R},\ \ \sigma=U{\sigma}_{R}U^\dag,
\]
and the transformations related to the Williamson decomposition are
\[
S=US_RU^\dag,\ \ D=UD_RU^\dag=D_R,\ \ K=Ui\Omega U^\dag.
\]
We explicitly write the connection between real and complex form of symplectic matrix,
\[
S_{R}=
\begin{bmatrix}
\alpha_R & \beta_R \\
\gamma_R & \delta_R
\end{bmatrix}=
\begin{bmatrix}
\operatorname{Re}\,[\A+\B] & -\operatorname{Im}\,[\A-\B] \\
\operatorname{Im}\,[\A+\B] & \operatorname{Re}\,[\A-\B]
\end{bmatrix}.
\]
Consequently, $\A$ and $\B$ needed for $R$ and $Q$ from Eq.~\eqref{eq:exact_multimode} can be expressed in the real form symplectic matrix elements as
\begin{subequations}
\begin{align}
\A&=\frac{1}{2}(\alpha_R+\delta_R+i\gamma_R-i\beta_R),\\
\B&=\frac{1}{2}(\alpha_R-\delta_R+i\gamma_R+i\beta_R).
\end{align}
\end{subequations}

Since all important matrices are related via this unitary transformation and traces and determinants are invariant under such transformations, it is clear that every formula we derived can be easily rewritten in the real form formalism by the formal substitution $\sigma\rightarrow\sigma_R$ and $K\rightarrow i\Omega$. On the other hand, the complex form provides a much more elegant structure and exposes the inner symmetries of symplectic and covariance matrices in more detail. Also, it is much easier to work with $K$ since it is diagonal, unitary, and Hermitian in contrast to non-diagonal skew-Hermitian matrix $\Omega$, providing much more convenient language.

\section{Pure states and the regularization}\label{app:pure_regularization}

In this section we derive a formula for the QFI for the states around the points of purity, i.e., for the points where $\lambda_i(\epsilon)=1$, but not necessarily $\lambda_i(\epsilon+d\epsilon)=1$. We can see that formula \eqref{GeneralQFI} is undefined at points $\lambda_i(\epsilon)=1$. This is the consequence that we assumed that for all $\Delta,\Lambda,\Gamma$ the Taylor expansion exist in a form given by Eqs.~\eqref{eqs:DeltaGammaLambda}, which is not true when $\lambda_i(\epsilon)=1$ for some $i\in\{1,2\}$.

For simplicity let us study the case where $\epsilon$ is such that the both $\lambda_1(\epsilon)=\lambda_2(\epsilon)=1$ and at the end of the calculation it will be clear how various cases work. For such states we have $D=I+D_1d\epsilon+D_2d\epsilon^2+O(d\epsilon^3)$. From Eqs.~\eqref{eqs:GDL_polished} we can see that $\Lambda=0$ and $\Delta=\Gamma$ and the Uhlmann fidelity~\eqref{eq:fidelity_basic_formula} simplifies to
\[\label{eq:fidelity_simplified}
\mathcal{F}\left(\rho_\epsilon,\rho_{\epsilon+d\epsilon}\right)=\frac{4}{\sqrt{\Delta}},
\]
where for simplicity we have omitted the part consisting of displacement. This coincides with the general formula~\cite{Spedalieri2013a} for fidelity between one pure and one pure or mixed Gaussian state, which is exactly our case. Now, because $\lambda_i(\epsilon)=1$ and the symplectic eigenvalue cannot fall below $1$, we have that either $\dot{\lambda}_i(\epsilon)=0$ or the derivative does not exist. If it exists and is, for example, positive, then for $d\epsilon<0$ we have $\lambda(\epsilon+d\epsilon)=1+\dot{\lambda}_i(\epsilon)d\epsilon+O(d\epsilon^2)<1$ for small enough $d\epsilon$, which cannot be true. From now on we assume that map $\sigma:\epsilon\longrightarrow\sigma(\epsilon)$ is differentiable, thus $D_1=0$ and $D=I+D_2d\epsilon^2+O(d\epsilon^3)$. The simplified formula \eqref{eq:fidelity_simplified} for fidelity gives $\mathcal{F}\left(\rho_\epsilon,\rho_{\epsilon+d\epsilon}\right)=1-\frac{1}{8}\big(\tr\big[P_1^2\big]-\tr\big[(KP_1)^2\big]+2\tr[D_2]\big)d\epsilon^2+O(d\epsilon^3)$, leading to the QFI,
\[\label{eq:QFI_almost_pure}
H(\epsilon)=\frac{1}{2}\Big(\tr\big[P_1^2\big]-\tr\big[(KP_1)^2\big]\Big)+\tr[D_2].
\]
Since $D_2=\frac{1}{2}\ddot{D}$, we have $\tr[D_2]=\sum_i\ddot{\lambda}_i$. We can rewrite the above equation in terms of $A=K\sigma$,
\[\label{eq:QFI_almost_pure2}
H(\epsilon)=\frac{1}{4}\big(2\tr\Big[A^{-1}\ddot{A}\big]-\tr\big[(A^{-1}\dot{A})^2\big]\Big).
\]
If the state remains pure with a small variation in $\epsilon$, then $\ddot{\lambda}_i=0$ and the above formula is equivalent to the pure state formula $H(\epsilon)=\frac{1}{4}\mathrm{tr}\big[(\sigma^{-1}\dot{{\sigma}})^2\big]$, which can be also derived from the equation for iso-thermal states~\eqref{eq:nu_pure}. We see that although general mixed-state formulae do not depend on the second derivative, formulae~\eqref{eq:QFI_almost_pure} and \eqref{eq:QFI_almost_pure2} do. But we can show that we achieve the same result from the mixed state formula when requiring the continuity of the QFI. If we assume that $\sigma:\epsilon\longrightarrow\sigma(\epsilon)\in C^{(2)}$, i.e., the second derivative exists everywhere and is continuous, then the QFI is continuous everywhere apart from the points where it is undefined. To make the QFI a continuous function we define the problematic points given by $\lambda_i(\epsilon)=1$ as limits
\[\label{def:problem_lambda}
\frac{\dot{\lambda}_i^2}{\lambda_i^2-1}(\epsilon):=\lim_{d\epsilon\rightarrow0}\frac{\dot{\lambda}_i(\epsilon+d\epsilon)^2}{\lambda_i(\epsilon+d\epsilon)^2-1}=\ddot{\lambda}_i(\epsilon),
\]
since $\lambda_i$ has expansion $\lambda_i(\epsilon+d\epsilon)=1+\frac{1}{2}\ddot{\lambda}_i(\epsilon)d\epsilon^2+O(d\epsilon^3)$. Now we see that the our definition of problematic values using the second derivative corresponds to the actual value of the QFI given by \eqref{eq:QFI_almost_pure}. On the other hand, note that authors of~\cite{Pinel2013b} choose rather $\frac{\dot{\lambda}_i(\epsilon)^2}{\lambda_i(\epsilon)^2-1}:=0$, which would correspond to using an exact formula for states which remain pure (setting $\ddot{\lambda}=0$ in \eqref{eq:QFI_almost_pure}), but may lead to discontinuities even for the smooth functions $\sigma(\epsilon)$. Nevertheless, to decide which convention is the right to choose, one should rather examine the validity of Cram\'er-Rao bound itself for such cases.

Similarly to definition \eqref{def:problem_lambda}, for the multi-mode formula \eqref{eq:exact_multimode} and $\lambda_i(\epsilon)=\lambda_j(\epsilon)=1$ we define
\[\label{def:problem_lambda_lambda}
\frac{(\lambda_i-\lambda_j)^2}{\lambda_i\lambda_j-1}(\epsilon):=
\lim_{d\epsilon\rightarrow0}\frac{(\lambda_i(\epsilon+d\epsilon)-\lambda_j(\epsilon+d\epsilon))^2}{\lambda_i(\epsilon+d\epsilon)\lambda_j(\epsilon+d\epsilon)-1}=0.
\]

Now we describe the regularization process which allows us to derive the correct value of the quantum Fisher information. In fact, all \eqref{GeneralQFISD},\eqref{GeneralQFI},\eqref{eq:Monras_polished_infty},\eqref{eq:multimode_P1} have problems when there exists at least one symplectic eigenvalue equal to one. The basic idea is that instead of computing $H(\epsilon):=H(\sigma(\epsilon))$ directly we calculate the quantum Fisher information for the regularized state $H(\nu\sigma(\epsilon))$, $\nu>1$ and at the end of the computation we perform the limit $\nu\rightarrow1$. This method must always give a finite result because by doing that we must arrive at an expression similar to Eq.~\eqref{eq:exact_multimode}. This becomes clear when we look at the derivation in~\ref{app:exact_multimode}. Also, the result it gives is certainly correct for any point where the formula is defined, we just need to make sure that this method gives the same definitions \eqref{def:problem_lambda},\eqref{def:problem_lambda_lambda} for the problematic points. For $\lambda_i(\epsilon)=\lambda_j(\epsilon)=1$ and differentiable $\sigma$ we have $\dot{\lambda}_i(\epsilon)=\dot{\lambda}_j(\epsilon)=0$ and
\begin{subequations}
\begin{align}
\lim_{\nu\rightarrow1}\frac{(\nu\dot{\lambda}_i(\epsilon))^2}{(\nu\lambda_i(\epsilon))^2-1}
&=\lim_{\nu\rightarrow1}\frac{0}{\nu^2-1}=0,\\
\lim_{\nu\rightarrow1}\frac{(\nu\lambda_i(\epsilon)-\nu\lambda_j(\epsilon))^2}{\nu\lambda_i(\epsilon)\nu\lambda_j(\epsilon)-1}
&=\lim_{\nu\rightarrow1}\frac{0}{\nu^2-1}=0.
\end{align}
\end{subequations}
Although the second equation gives the same definition as required by Eq.~\eqref{def:problem_lambda_lambda}, we can see that the first equation does not give the same definition as Eq.~\eqref{def:problem_lambda} because the problematic term here is set to zero by the regularization procedure. As a result, for every $\epsilon$ such that $\lambda_i(\epsilon)=1$ we need to add additional $\ddot{\lambda}_i(\epsilon)$ to the regularized version of the quantum Fisher information. Therefore, we have shown that we can calculate the quantum Fisher information at the points of purity as a limit
\[
H(\epsilon)=\lim_{\nu\rightarrow1}H\big(\nu\sigma(\epsilon)\big)+\!\!\!\!\!\!\sum_{i:\lambda_i(\epsilon)=1}\!\!\!\!\!\!\ddot{\lambda}_i(\epsilon).
\]

\section{Estimation of the remainder in the multi-mode formula}\label{app:Remainder}
Here we prove the bound on the remainder of the general multi-mode formula. We consider the Williamson decomposition $\sigma=SDS^\dag$. An element of the sum Eq.~\eqref{eq:Monras_polished} can be written as
\[
a_n=\tr\big[A^{-n}\dot{A}A^{-n}\dot{A}\big]=\tr\big[{D^{-n}BD^{-n}B}\big],
\]
where $B=S^\dag\dot{A}(S^\dag)^{-1}K^{-n-1}$. We can derive the inequalities
\[\label{eq:Remainder1}
\begin{split}
\norm{a_n}&=\norm{\sum_{k,l}\frac{1}{\lambda_k^n\lambda_l^n}B_{kl}B_{lk}}
\leq\norm{\sum_{k,l}\frac{1}{\lambda_k^{n}\lambda_l^{n}}\norm{B_{kl}}\norm{B_{lk}}}\\
&\leq\frac{1}{\lambda_{\mathrm{min}}^{2n}}\norm{\sum_{k,l}\norm{B_{kl}}\norm{B_{lk}}}\leq\frac{1}{\lambda_{\mathrm{min}}^{2n}}\bigg(\sum_{k,l}\norm{B_{kl}}^2\bigg)\\
&=\frac{1}{\lambda_{\mathrm{min}}^{2n}}\big(\tr[B^\dag B]\big),
\end{split}
\]
where the last inequality is the Cauchy-Schwarz inequality between $B_{ij}$ and $B_{ji}$ considered as vectors with $4N^2$ entries where $N$ is number of modes, $\lambda_{\mathrm{min}}:=\min_{i}\{\lambda_i\}$ is the smallest symplectic eigenvalue. Defining the Hermitian matrix $C:=S^\dag \dot{A}KS$ we have
\[\label{eq:Remainder2}
\begin{split}
\tr[(A\dot{A})^2]&=\tr[C^\dag D C D]=\sum_{k,l}\norm{C_{kl}}^2\lambda_k\lambda_l\\
&\geq\lambda_{\mathrm{min}}^2\tr[C^\dag C]=\lambda_{\mathrm{min}}^2\tr[B^\dag B].
\end{split}
\]
Combining \eqref{eq:Remainder1} and \eqref{eq:Remainder2} gives
\[
\norm{a_n}\leq\tr\big[(A\dot{A})^2\big]\lambda_{\mathrm{min}}^{-2n-2}.
\]
For $\lambda_{\mathrm{min}}>1$ we can estimate the remainder,
\[
\norm{R_M}\leq \frac{\tr\big[(A\dot{A})^2\big]}{2}\!\!\sum_{n=M+1}^\infty\!\!\!\!{\lambda_{\mathrm{min}}^{-2n-2}}
=\frac{\tr\big[(A\dot{A})^2\big]}{2\lambda_{\mathrm{min}}^{2M+2}(\lambda_{\mathrm{min}}^2-1)}.
\]

\section{Derivation of the exact formula for a multi-mode Gaussian state.}\label{app:exact_multimode}
Here we derive the general formula for the QFI given by Eq.~\eqref{eq:exact_multimode}. Denoting $D_{0ii}:=(D_0)_{ii}$, we start by evaluating an infinite sum from the equation \eqref{eq:multimode_P1}:
\[
\begin{split}
&\sum_{n=1}^\infty\tr\Big[\big(D_0^2-I\big)\big(D_0^{-n}K^{-n}P_1\big)^2\Big]\\
&=\!\!\sum_{n=1}^\infty\!\tr\Big[\big(\!D_0^2-I\!\big)\big(\!D_0^{-2n+1}KP_1\!\big)^2\Big]\!+\!\tr\Big[\big(\!D_0^2-I\!\big)\big(\!D_0^{-2n}P_1\!\big)^2\Big]\\
&=\!\sum_{i,j}\big(D_{0ii}^2-1\big)\bigg(\sum_{n=1}^\infty(D_{0ii}D_{0jj})^{-2n+1}\bigg)(KP_1)_{ij}(KP_1)_{ji}\\
&\ \ \ \ +\big(D_{0ii}^2-1\big)\bigg(\sum_{n=1}^\infty(D_{0ii}D_{0jj})^{-2n}\bigg)(P_1)_{ij}(P_1)_{ji}\\
&=\!\!\sum_{i,j}\!\!\frac{D_{0ii}^2-1}{D_{0ii}^2D_{0jj}^2-1}\Big((\!KP_1\!)_{ij}(\!KP_1\!)_{ji}\!+\!D_{0ii}D_{0jj}(\!P_1\!)_{ij}(\!P_1\!)_{ji}\Big).
\end{split}
\]
We combine this expression with the first part $\tr[P_1^2]$ in \eqref{eq:multimode_P1}:
\[
\begin{split}
&\tr\big[P_1^2\big]-\sum_{n=1}^\infty(\cdot)
=\sum_{i,j}\frac{D_{0ii}^2\big(D_{0jj}^2-1\big)}{D_{0ii}^2D_{0jj}^2-1}(P_1)_{ij}(P_1)_{ji}\\
&\ \ \ \ -\frac{D_{0ii}D_{0jj}\big(D_{0ii}^2-1\big)}{D_{0ii}^2D_{0jj}^2-1}(KP_1)_{ij}(KP_1)_{ji}.
\end{split}
\]
Now we use definitions \eqref{def:S_and_D},\eqref{def:S_and_P}, $P_1=S^{-1}\dot{S}=KS^\dag K\dot{S}$, $K=I\oplus-I$ and $D_0=\mathrm{diag}(\lambda_1,\dots,\lambda_N,\lambda_1,\dots,\lambda_N)$ to rewrite the expression in terms of the symplectic eigenvalues and sub-matrices $R$ and $Q$,
\[
\begin{split}
&\tr\big[P_1^2\big]\!-\!\!\sum_{n=1}^\infty(\cdot)\!=\!\!\sum_{i,j=1}^N\!2\frac{\lambda_i^2(\lambda_j^2\!-\!1)\!-\!\lambda_i\lambda_j(\lambda_i^2\!-\!1)}{\lambda_i^2\lambda_j^2-1}\operatorname{Re}(R_{ij}R_{ji})\\
&\ \ \ \ \ \ +\!2\frac{\lambda_i^2(\lambda_j^2-1)+\lambda_i\lambda_j(\lambda_i^2-1)}{\lambda_i^2\lambda_j^2-1}\operatorname{Re}(\overline{Q}_{ij}Q_{ji})\\
&=\sum_{i,j=1}^N\!2\frac{\lambda_i(\lambda_j\!-\!\lambda_i)}{\lambda_i\lambda_j-1}\operatorname{Re}(R_{ij}R_{ji})\!+\!2\frac{\lambda_i(\lambda_j\!+\!\lambda_i)}{\lambda_i\lambda_j+1}\operatorname{Re}(\overline{Q}_{ij}Q_{ji}).
\end{split}
\]
$\operatorname{Re}(\overline{Q}_{ij}Q_{ji})$ and $\operatorname{Re}(R_{ij}R_{ji})$ are both symmetric. This is why we decompose parts consisting of symplectic eigenvalues into its symmetric and antisymmetric parts: $2\frac{\lambda_i(\lambda_j\!-\!\lambda_i)}{\lambda_i\lambda_j-1}=-\frac{(\lambda_j-\lambda_i)^2}{\lambda_i\lambda_j-1}+\frac{\lambda_j^2-\lambda_i^2}{\lambda_i\lambda_j-1}$, $2\frac{\lambda_i(\lambda_j\!+\!\lambda_i)}{\lambda_i\lambda_j+1}=\frac{(\lambda_j+\lambda_i)^2}{\lambda_i\lambda_j+1}+\frac{\lambda_i^2-\lambda_j^2}{\lambda_i\lambda_j+1}$ and the antisymmetric part vanishes under the sum. In addition, from identity \eqref{id:P_1_symplectic} we have $R^\dag=-R$ and $Q^{T}=Q$ giving us $\operatorname{Re}(R_{ij}R_{ji})=-\norm{R_{ij}}^2$, $\operatorname{Re}(\overline{Q}_{ij}Q_{ji})=\norm{Q_{ij}}^2$. The diagonal part is simply $\frac{1}{2}\tr\big[(D_0+K)^{-1}D_0^{-1}D_1^2\big]=\sum_{i=1}^N\frac{\dot{\lambda_i}^2}{\lambda_i^2-1}$. Combining all these yields
\[
\begin{split}
H(\epsilon)&=\sum_{i,j=1}^N\frac{(\lambda_i-\lambda_j)^2}{\lambda_i\lambda_j-1}\norm{R_{ij}}^2+\frac{(\lambda_i+\lambda_j)^2}{\lambda_i\lambda_j+1}\norm{Q_{ij}}^2\\
&+\sum_{i=1}^N\frac{\dot{\lambda_i}^2}{\lambda_i^2-1}+2\dot{\bd}^\dag\sigma^{-1}\dot{\bd}.
\end{split}
\]
The definitions of expressions for problematic points $\lambda_i(\epsilon)=\lambda_j(\epsilon)=1$ come from the assumption of the continuity of the QFI and correspond to the knowledge gained in~\ref{app:pure_regularization}.

\section{Calculating the example}\label{app:example}
Here, to illustrate how the general formula~\eqref{eq:exact_multimode} works, we calculate the quantum Fisher information for an example of a one-mode Gaussian state. For $N=1$ the general formula~\eqref{eq:exact_multimode} has a compact form
\[\label{eq:exact_onemode}
H(\epsilon)=4\frac{\lambda_1^2}{\lambda_1^2+1}\norm{Q_{11}}^2+\frac{\dot{\lambda_1}^2}{\lambda_1^2-1}+2\dot{\bd}^\dag\sigma^{-1}\dot{\bd},
\]
where $Q_{11}=\ov{\A}\dot{\B}-\B\dot{\ov{\A}}$.

We consider a task of estimating a squeezing parameter of a squeezing channel, with a squeezed rotated displaced thermal state as an input state. In the complex form of the covariance matrix formalism, the one-mode squeezing operator with a squeezing parameter $r$, and the rotation operator via angle $\theta$ are
\[\label{def:squeezing_matrix}
S_r=\begin{bmatrix}
\cosh r & -\sinh r \\
-\sinh r & \cosh r
\end{bmatrix},\quad
R_\theta=\begin{bmatrix}
e^{-i\theta} & 0 \\
0 & e^{+i\theta}
\end{bmatrix}.
\]
First, we create an input state by squeezing, rotating, and displacing a thermal state. We obtain a state given by its first and the second moments,
\[
\bd_0,\quad \sigma_0=R_\theta S_rDS_r^\dag R_\theta^\dag,
\]
where $\bd_0=(\tilde{d_0},\ov{{\tilde{d_0}}})^T$ is the initial displacement, $\sigma_0$ the initial covariance matrix, and $D=\mathrm{diag}(\lambda_1,\lambda_1)$ is the covariance matrix of a thermal state. For a harmonic oscillator with frequency $\omega$ we have $\lambda_1=\coth(\frac{\omega\hbar}{2kT})$.
Now we feed the prepared state into the channel which we consider to be again a simple squeezing operation with the unknown squeezing parameter $\epsilon$ we want to estimate. We obtain the final state,
\[
\bd=S_\epsilon \bd_0,\quad \sigma=S_\epsilon R_\theta S_rDS_r^\dag R_\theta^\dag S_\epsilon^\dag.
\]
It is clear from the construction that the diagonalizing symplectic matrix needed to compute $Q_{11}$ in formula~\eqref{eq:exact_onemode} is $S=S_\epsilon R_\theta S_r$. Considering definitions~\eqref{def:S_and_D} and~\eqref{def:squeezing_matrix} we derive
\begin{subequations}
\begin{align}
\A&=e^{-i\theta} \cosh\epsilon\cosh r+e^{+i\theta} \sinh \epsilon\sinh r,\\
\B&=e^{-i\theta} \cosh\epsilon\sinh r+e^{+i\theta} \sinh \epsilon\cosh r.
\end{align}
\end{subequations}
For the part containing displacement we need to compute the inverse of the covariance matrix, $\sigma^{-1}=KSD^{-1}S^\dag K$, and the change of displacement, $\dot{\bd}=\dot{S}_\epsilon \bd_0$. The middle part of expression~\eqref{eq:exact_onemode} is simply zero, because $\lambda_1$ is independent of $\epsilon$. In other words, the purity of the system remains the same. Now we have everything prepared to calculate the quantum Fisher information,
\[
\begin{split}
H(\epsilon)\!&=\!\frac{4\lambda_1^2}{\lambda_1^2+1}\big(\cosh^4r+\sinh^4r-2\cos(4\theta)\cosh^2r\sinh^2r\big)\\
&+\frac{4\norm{\tilde{d_0}}^2}{\lambda_1}\big(\cosh(2r)+\cos(2(\theta-\phi))\sinh(2r)\big).
\end{split}
\]

\section{Table of frequently used notation}\label{app:notations}
\begin{tabular}{ r | l }
$K$ & A constant matrix defining \\
& a complex represenation of the real symplectic\\
& group defined in Eq.~\eqref{def:commutation_relation}. \\
$\bd$ & The displacement vector, Eq.~\eqref{def:covariance_matrix}.\\
$\sigma$ & The covariance matrix, Eq.~\eqref{def:covariance_matrix}.\\
$A$ & $A:=K\sigma$, a multiple of the two previously\\
& mentioned matrices.\\
$S$ & The symplectic matrix from the Williamson\\
& decomposition of $\sigma$, Eq.~\eqref{def:S_and_D}.\\
$\alpha, \beta$ & Submatrices of the matrix $S$, Eq.~\eqref{def:S_and_D}.\\
$\!\!\!D\!,\!D_{\!0},\!L$ & The diagonal matrices consisting of\\
& the symplectic eigenvalues, Eqs.~\eqref{def:S_and_D},\eqref{eqn:matrix_not}.\\
$\dot{~}$ & The derivative with respect to the parameter\\
& we want to estimate.\\
$P_1$ & $P_1:=S^{-1}\dot{S}$, an element of the Lie algebra\\
&associated with the symplectic group,\eqref{id:P_1_symplectic},\eqref{def:S_and_P}.\\
$R, Q$ & Submatrices of the matrix $P_1$, Eq.~\eqref{def:S_and_P}.\\
$\overline{\phantom{\alpha}}$, $^T$, $^\dag$  & Complex conjugate, transpose,\\
& conjugate transpose respectively.\\
\end{tabular}
\linebreak
\bibliographystyle{apsrev}
\bibliography{QFI_multimode_GS_BiBTeX}

\end{document}